\documentclass[a4paper,oneside,11pt,reqno]{amsart}
\usepackage{amssymb,bbm}
\usepackage{amsthm}
\usepackage{amsmath}
\usepackage{cite}
\usepackage[margin=2.9cm]{geometry} 
\usepackage[foot]{amsaddr}

\usepackage{hyperref}
\hypersetup{
	pdftitle = {Simplified exact SICs}, 
	pdfauthor = {Marcus Appleby, Ingemar Bengtsson}
}

\expandafter\def\expandafter\normalsize\expandafter{%
    \normalsize
    \setlength\abovedisplayskip{12pt}
    \setlength\belowdisplayskip{12pt}
    \setlength\abovedisplayshortskip{10pt}
    \setlength\belowdisplayshortskip{10pt}
}

\newcommand{\one}{{\rm 1\hspace*{-0.4ex}\rule{0.1ex}{1.52ex}\hspace*{0.2ex}}}
\makeatletter
\renewcommand{\@secnumfont}{\bfseries}
\renewcommand\section{\@startsection{section}{1}%
 \z@{.7\linespacing\@plus\linespacing}{.5\linespacing}%
  {\indent\normalfont\bfseries}}
\makeatother

\begin{document}

\vspace{6cm}

\begin{center}

{\Large SIMPLIFIED EXACT SICS}\footnote{Resubmitted on May 23, 2019.}

\vspace{14mm}

{\large Marcus Appleby}$^{*}$

\

{\large Ingemar Bengtsson}$^{\dagger}$

\vspace{14mm}

$^*${\small {\sl  Centre for Engineered Quantum Systems, School of Physics, \\
University of Sydney, Australia}}

\

$^{\dagger}${\small {\sl Stockholms Universitet, AlbaNova, Fysikum, \\
Stockholm, Sverige}}

\vspace{14mm}

{\bf Abstract:}

\vspace{7mm} 

\parbox{112mm}{
\noindent In the standard basis exact expressions for the components of SIC 
vectors (belonging to a symmetric informationally complete POVM) are typically 
very complicated. We show that a simple transformation to a basis adapted to the 
symmetries of a fiducial SIC vector can result in a massive reduction in complexity. 
We rely on a conjectural number theoretic connection between SICs in dimension 
$d_j$ and SICs in dimension $d_{j+1} = d_j(d_j-2)$. We focus on the sequence 5, 
15, 195, \dots  
We rewrite Zauner's exact solution for the SIC in dimension 5 to make its 
simplicity manifest, and use our adapted basis to convert numerical solutions 
in dimensions 15 and 195 to exact solutions. Comparing to the known exact 
solutions in dimension 15 we find that the simplification achieved is dramatic. 
The proof that the exact vectors are indeed SIC fiducial vectors, also in 
dimension 195, is guided by the standard ray class hypothesis about the algebraic 
number fields generated by the SICs. In the course of the calculation we introduce 
SIC adapted generators for the ray class field. We conjecture 
that our result generalizes to every dimension in the particular sequence we consider.} 
\end{center}

\newpage

\section{Introduction}\label{sec:intro}

\vspace{5mm}

\noindent A SIC is a delicately placed collection of $d^2$ unit vectors in ${\bf C}^d$. 
The definition requires that the vectors give rise to a resolution of the identity, and that 
all scalar products between distinct vectors have the same absolute value. 
Among alternative names 
``maximal equiangular tight frame'' is more descriptive, ``SIC-POVM'' less so. 
The question whether SICs exist in all finite dimensions is of foundational interest 
in quantum theory, and it arises as a kind of engineering problem in classical 
signal processing \cite{FHS}. 
Twenty years ago Zauner conjectured that SICs exist in every dimension, that they 
form orbits under the finite Weyl--Heisenberg group, and that they enjoy certain 
symmetries derived from the unitary automorphism group of the latter \cite{Zauner}. 
Until recently progress in the area consisted in finding numerical and exact solutions, 
and studying their properties \cite{Renes, Marcus, Scott, Zhu, Andrew}. It then 
emerged that the components of the SIC vectors, relative to the standard basis singled
out by the group, generate algebraic number fields of considerable pure mathematical 
interest \cite{AYAZ, AFMY}. As a result there is a new motivation for studying SICs, 
especially since they may establish a link between physics 
and an area of pure mathematics that so far has not interacted with physics. 

Since it forms an orbit under a group, a SIC can be presented by listing the components 
of a single fiducial vector. The highest dimension for which an exact solution has 
been published is $323 = 19\cdot 17$, with $124$ taking second place \cite{GS}. 
In unpublished work Grassl and Scott have pushed the record to $1299$, with 
$2208 = 48\cdot 46$ being the record dimension for a numerical solution 
\cite{SGprivate}. The catch 
is that the actual numbers look complicated. In high dimensions they look ghastly. 
In this paper we will perform a simple transformation from the standard basis to 
a basis that is adapted to the (known or conjectured) symmetries of the SIC fiducial 
vectors \cite{DMA}. We then consider a number theoretically connected 
sequence of dimensions, and show that for the first three dimensions in this sequence 
(namely dimensions 5, 15, and 195) there exist SIC fiducials such that all their 
components relative to the adapted basis are of the form 

\begin{equation} (q_1+q_2\sqrt{3})^{\frac{1}{2}}
(q_3+iq_4)^{q_5} (q_6+iq_7\sqrt{3})^{q_8}e^{2\pi i q_9} \ , \label{comp} \end{equation}

\noindent where $q_1, \dots , q_9$ are rational numbers with modest denominators. 
The full expressions for 
the components of fiducial 195d are given in Table \ref{tab:modulid} below 
supplemented by eqs.\ (\ref{pyt1}) to (\ref{pyt4}). This massive reduction of 
the expected complexity is brought about by adapting the mode of presentation to 
the SIC, to the best of our ability. We conjecture that similarly simplified 
solutions occur in an infinite sequence of dimensions. 

The reason for seeking a simpler way of writing SICs  is not simply a matter of 
aesthetics, or the desire to save space. It is related to the surprising link between 
the SIC existence problem on the one hand, and major unsolved problems in algebraic 
number theory on the other \cite{AFMY}. To see what is involved, it is interesting to 
compare SICs to complete sets of mutually unbiased bases, another special configuration 
of vectors of interest in quantum information theory \cite{Ivanovic}. At first 
glance the standard expressions for the vectors in the latter are much simpler than 
the expressions for a SIC in the same dimension. However, as observed in an appendix 
of ref. \cite{ABDF}, that is only because we are not comparing like with like. The SICs 
are expressed in terms of radicals. The mutually unbiased bases are expressed in terms of 
roots of unity, which in their turn are expressed using the exponential function. If 
one writes the roots of unity in terms of radicals then the expressions can become equally 
complicated. This suggests that the problem is to find functions which play the same 
role for the SIC numbers that the exponential function does for the roots of unity. If 
one could do that, and if one could find appropriate identities 
satisfied by these functions, then one might be able to solve the SIC existence 
problem and, incidentally, one may have taken a significant step in the direction of 
solving Hilbert's $12^{\rm{th}}$ problem. Recently Kopp has made a specific proposal 
along these lines: namely that in prime dimensions equal to 2 modulo 3 ray class $L$-functions 
play such a role \cite{Kopp}.  The result we prove establishes a different 
kind of simplification, using Hilbert space geometry rather than number theory 
as a guide. We hope it will prove useful too. 

The first link between SICs and algebraic number theory was the observation \cite{AYAZ} 
that (ratios between) the components of the SIC vectors in dimension $d\geq 4$, 
when expressed in the basis singled out by the Weyl--Heisenberg group, lie in number 
fields that contain the real quadratic field ${\mathbb Q}(\sqrt{D})$, where 

\begin{equation} D = \ \mbox{the square-free part of} \ (d+1)(d-3) \ . \end{equation}

\noindent Given an integer 
$D \geq 2$ there is an infinite number of integers $d$ leading to this value of 
$D$. What is more, given $D$ and $d$ there exists a ray class field with 
base field ${\mathbb Q}(\sqrt{D})$ and conductor $d$. In fact there exists four 
such ray class fields, differing in whether they are ramified in no, one, or two 
infinite places. The distinction is important, but for our purposes it is the 
ray class field ramified at two infinite places that we need, and this is unique 
given $D$ and given the conductor. From now on, when we refer to ``the ray class 
field'' we mean the one ramified at both infinite places (which is also the largest 
one), and we refer elsewhere for an elucidation of this point \cite{AFMY}. For a 
long time algebraic number theorists have been concerned with trying to obtain a 
better understanding of these ray class fields. Remarkably, in every dimension 
$d$ so far investigated, the (ratios between the) components of a fiducial 
vector for a special SIC, together with appropriate roots of unity, generate 
the ray class field with conductor $d$, or $2d$ if $d$ is even \cite{AFMY}. 
The Ray Class Hypothesis states 
that this happens in all dimensions, so that the number fields containing these SICs 
are conjecturally known. To avoid any misunderstanding, the ray class fields typically 
have much higher degrees than their base fields. If there are several inequivalent 
SICs, the others lie in fields containing the ray class field \cite{ACFW}.

We can now see how the dimension ladders arise. Make the substitution 

\begin{equation} d \rightarrow d(d-2) \hspace{5mm} \Rightarrow \hspace{5mm} 
(d+1)(d-3) \rightarrow (d-1)^2(d+1)(d-3) \ . \end{equation}

\noindent The square-free part $D$, and hence the base field of the SIC, is 
unchanged. What is more, a ray class field 
is contained in another if its conductor divides that of the other. This gives 
rise to infinite sequences of dimensions $d_j$, defined by the recursion relation 
$d_{j+1} = d_j(d_j-2)$, for which the ray class hypothesis implies number theoretic 
connections between SICs in different dimensions. These connections are reflected 
by geometrical properties of the SICs \cite{GS, ABDF}, and raise the hope that 
closed form expressions for SICs can be found in infinite sequences of dimensions. 
However, here we climb only two rungs up the ladder that starts with $d =5$ and $D = 3$. 

A curious feature of our simplified solutions is that the components (\ref{comp}) 
of the SIC fiducial vectors relative to the normalized adapted basis typically 
lie in quadratic extensions of the minimal number field. On closer inspection one 
finds that this feature can be regarded as an artefact of the normalization of 
the basis vectors, so that the ray class hypothesis is vindicated. 

We have organized our paper as follows. The solutions are given in Section 
\ref{sec:solutions}. Section \ref{sec:groups} is a quick review of known facts about 
the relevant groups, and the adapted bases are computed in Section \ref{sec:basis}. 
Section \ref{sec:numerics} explains how, in favourable cases, the adapted bases 
can be used to convert numerical fiducial vectors to exact ones using only modest 
precision. In Section \ref{sec:checks} we explain the lengthy calculations we use 
to prove that the candidate solutions from Section \ref{sec:solutions} are indeed SIC 
fiducials. These calculations depend heavily on the ray class hypothesis. In the 
concluding Section \ref{sec:summary} we set our solutions in context.

\vspace{1cm}

\section{Solutions}\label{sec:solutions}

\vspace{5mm}

\noindent We present the matter back to front, that is we first give solutions for 
SIC fiducials, and afterwards define the bases in which they are expressed.  
The solution in dimension 5 is taken from Zauner's thesis \cite{Zauner}. It is 

\begin{equation} |\Psi_{\bf 0}\rangle = \sqrt{p_0}|e^{(5)}_0\rangle + 
\sqrt{p_1}(P_5)^\frac{1}{4}|e^{(5)}_1\rangle \ , \label{d5a} \end{equation}

\noindent where 

\begin{equation} p_0 = \frac{3 - \sqrt{3}}{4} \ , \hspace{8mm} p_1 = 
\frac{1 + \sqrt{3}}{4} \ , \end{equation}

\begin{equation} P_5 = - \frac{3}{5} + \frac{4i}{5} \ . \end{equation}

\noindent The vectors $|e^{(5)}_0\rangle$ and $|e^{(5)}_1\rangle$ are eigenvectors 
of a unitary operator connected to the SIC. They span what is known as the Zauner 
subspace, and will be easy to describe in words once we have presented the background 
(in Section \ref{sec:groups}). The form of the components of the vector is interesting. It is 
remarkable that the squares of their absolute values ($p_0$ and $p_1$) lie in the 
base field ${\mathbb Q}(\sqrt{3})$. Moreover the phase factor is the fourth root of a 
phase factor formed from a Pythagorean triple of integers. That is, $P_5$ is a 
phase factor because 

\begin{equation} 3^2 + 4^2 = 5^2 \ . \end{equation}

\noindent This relation was remarked upon in ancient Egypt. 

Group theoretically preferred adapted bases for the Zauner subspaces exist also in 
the higher dimensions considered next. They will be explicitly constructed 
in Section 4. For now, suffice it to say that they can be calculated by 
hand in all the cases we consider. Given these adapted bases we expand the SIC 
fiducial as

\begin{equation} |\Psi_{\bf 0}\rangle = \sum_{r=0}^{d_S-1} \sqrt{p_r}e^{i\nu_r}
|e^{(d)}_r\rangle \ , \label{15fid} \end{equation}

\noindent where $d$ is the dimension and $d_S$ is the dimension of the subspace 
(singled out by its symmetries) in which 
the fiducials live. In the Tables we give the exact solutions 
for the squares $p_r$ of the absolute values and for the phases $e^{i\nu_r}$. 
The normalization of the vector is ensured by 

\begin{equation} \sum_r p_r = 1 \ . \end{equation}

\noindent An overall phase factor is fixed by the requirement that $e^{i\nu_0} = 1$. 

In dimension $15 = 5\cdot 3$ there exist four unitarily and anti-unitarily 
inequivalent SICs. The ones we are interested in are labelled 15d and 15b \cite{Scott}. 
15d is the ray class SIC. It has higher symmetry and sits in a four dimensional subspace 
of the six dimensional Zauner subspace. In dimension 195 four SICs are available (see our 
Acknowledgements) in numerical form. The ones we are interested in are 195d, 
which is aligned to 15d in a sense that 
can be made precise, and 195b which is similarly aligned to 15b \cite{ABDF}. 
They both live in a 36 dimensional subspace of the Zauner subspace. 195d has higher 
symmetry and sits in a 19 dimensional subspace. 

As we move up the dimension ladder new Pythagorean triple phase factors enter, including 
some of a new kind constructed using also $\sqrt{3}$ in the numerator. It remains 
true that all the phase factors occurring in the dimensions 15 and 195 are triple phase 
factors multiplied with roots of unity. Since the triples enjoy a group structure, 
so that those with composite numbers in their denominators can be expressed as products 
of ones with prime denominators, we have expressed everything in terms of 

\begin{equation} \hspace{5mm} P_5=-\frac{3}{5}+\frac{4i}{5} \hspace{20mm}  
Q_2=-\frac{1}{2}+\frac{i\sqrt{3}}{2} = \omega_3 \ 
\label{pyt1} \end{equation}
 
\begin{equation} P_{13}=-\frac{5}{13}+\frac{12i}{13} \hspace{18mm} 
Q_{13}=-\frac{11}{13}+\frac{4i\sqrt{3}}{13} \ \ \label{pyt2} \end{equation}

\begin{equation} P_{37} = - \frac{12}{37} + \frac{35i}{37} \hspace{18mm} 
 Q_{37} = - \frac{13}{37} + \frac{20 i \sqrt{3}}{37} \ \ \label{pyt3} \end{equation}

\begin{equation} P_{241} = - \frac{120}{241} + \frac{209i}{241} \hspace{18mm} 
Q_{241} = - \frac{143}{241} + \frac{112i\sqrt{3}}{241} \ . \label{pyt4} \end{equation}

\noindent It was known to Diophantus that the primes occurring in the denominators 
of the triples $P_n$ are necessarily equal to 1 mod 4. One way to see this is 
that the prime in the denominator must split over ${\mathbb Q}(i)$ if the expression 
is to be a phase factor. For the $Q_n$ with $n > 2$ the prime in the denominator must 
split over ${\mathbb Q}(\omega_3)$, which forces the prime to equal 1 mod 3  \cite{HW}. 
The exceptional case $Q_2 = \omega_3$ is an algebraic unit. 

\begin{table}[b]
\caption{{\small Moduli squared and relative phases in two $d = 15$ SICs. }}
  \smallskip
{\renewcommand{\arraystretch}{1.4}
\begin{tabular}
{|ll||ll|}\hline \hline
15d & & 15b & \\
(Moduli)$^2$ & Phases & (Moduli)$^2$ & Phases \\
\hline			
$p_0 = \frac{\sqrt{3}}{4}$ & $e^{i\nu_0} = 1 $ & 
$p_0 = \frac{\sqrt{3}}{8}$ & $e^{i\nu_0} = 1 $ \\ 
$p_1 = \frac{2-\sqrt{3}}{4}$ & $e^{i\nu_1} = -i \left( P_5\right)^{\frac{1}{4}}$& 
$p_1 = \frac{4-\sqrt{3}}{8}$ & 
$e^{i\nu_1} = -i (P_5)^{-\frac{1}{4}}\left( - Q_{13}P_{13}\right)^\frac{1}{4}$ \\ 
$p_2=\frac{\sqrt{3}}{8}$ 
& $e^{i\nu_2} =\omega_{12} \left( P_5 \right)^{\frac{1}{3}}$ &
$p_2=\frac{2\sqrt{3}-3}{8}$ 
& $e^{i\nu_2} = \omega_{12}^5\left( P_5 \right)^{-\frac{1}{3}}$ \\ 
$p_3=\frac{4-\sqrt{3}}{8}$ & 
$e^{i\nu_3} =\omega_3 (P_5)^\frac{1}{12}
\left( - Q_{13}P_{13}\right)^{-\frac{1}{4}}$ &
$p_3= \frac{1}{8} $ & $e^{i\nu_3} =\omega_{12}^{11}\left( P_5\right)^{-\frac{1}{12}}$ \\ 
& & $p_4= \frac{6-3\sqrt{3}}{8}$ & $e^{i\nu_4} = - 1$ \\ 
& & $p_5=\frac{\sqrt{3}}{8}$ & $e^{i\nu_5} =\left( P_5\right)^{-\frac{1}{4}}$ \\ 
\hline \hline
\end{tabular}
}
\label{tab:modulidb}
\end{table} 

The absolute values squared and the phases needed in eq.\ (\ref{15fid}) are given in 
Table \ref{tab:modulidb} for 15db and Table \ref{tab:modulid} for 195d. The candidate 
solution for 195b is placed in an Appendix as Table \ref{tab:modulib}, because for 195b 
we have not carried out the entire proof that the candidate solution is indeed an exact 
solution. We observe that a Pythagorean triple with 37 in the denominator occurs as a 
phase factor only in fiducial 195d, while a triple with 241 in the denominator occurs 
only in 195b. The simplicity of the components when expanded in 
our adapted basis should be judged by means of a comparison to how fiducial 15d looks 
when expanded in the standard basis \cite{Scott}.

\begin{table}[h]
\caption{{\small Moduli squared and relative phases in 195d.}}
  \smallskip
\hskip 0.2cm
{\renewcommand{\arraystretch}{1.6}
\begin{tabular}
{|c|l|l}\hline \hline
$p_i$ 
& Phases $e^{i\nu_i}$  \\
\hline			
$p_0 = \frac{12\sqrt{3}-9}{182}$
& $e^{i\nu_0} = 1$  \\
$p_1=\frac{19-8\sqrt{3}}{182}$
& $e^{i\nu_1} = (P_5)^\frac{1}{4}\left( - Q_{13}P_{13}\right)^{\frac{1}{4}}$ \\ 
$p_2 = \frac{12\sqrt{3}-9}{364}$
& $e^{i\nu_2} = \omega_3^2\left( P_5\right)^\frac{1}{3}$\\ 
$p_3 = \frac{41-20\sqrt{3}}{364}$
& $e^{i\nu_3} = 
- \omega_{24}(P_5)^\frac{1}{12} \left( i P_{37}Q_{37}\right)^\frac{1}{4}$ \\
$p_4 = \frac{55\sqrt{3}-90}{546}$
& $e^{i\nu_4} = -i\omega_{24} \left( P_5\right)^{-\frac{1}{2}}$ \\ 
$p_5 = \frac{20-5\sqrt{3}}{182}$ 
& $e^{i\nu_5} = \omega_{24} \left( P_5\right)^\frac{3}{4}$ \\ 
$p_6 = \frac{4\sqrt{3}-3}{273}$ 
& $e^{i\nu_6} = \omega_3^2\left( P_5\right)^\frac{1}{3}$ \\  
$p_7 = \frac{19-8\sqrt{3}}{91}$
&  $e^{i\nu_7} = \omega_{12}^2
(P_5)^\frac{1}{12}\left( -Q_{13}P_{13}\right)^{-\frac{1}{4}}$ \\ 
$p_8 = \frac{5-2\sqrt{3}}{14}$
& $e^{i\nu_8} = \omega_{12}^{10} (P_5)^\frac{1}{3} 
\left( - \frac{\omega_3Q_{13}}{P_{13}}\right)^\frac{1}{12}$ \\ 
$p_9 = \frac{2\sqrt{3}-3}{14}$
& $e^{i\nu_9} = \omega_{12}^2 (P_5)^\frac{1}{12}
\left( -\frac{P_{13}}{\omega_3Q_{13}}\right)^\frac{1}{6}$ \\ 
$p_{10} = \frac{2\sqrt{3}-3}{21}$
& $e^{i\nu_{10}} = \omega_{12}^7 (P_5)^\frac{1}{6}
\left( - \frac{P_{13}}{\omega_3Q_{13}}\right)^\frac{1}{12}$ \\ 
$p_{11} = \frac{\sqrt{3}}{21}$ 
& $e^{i\nu_{11}} = \omega_{24} \omega_{12}^2 
\left( - \frac{P_{13}}{\omega_3Q_{13}}\right)^\frac{1}{12}$ \\ 
$p_{12} = \frac{2-\sqrt{3}}{7}$
& $e^{i\nu_{12}} = \omega_{24} \omega_3^2 (P_5)^\frac{1}{4}
\left( - \frac{P_{13}}{\omega_3Q_{13}}\right)^\frac{1}{12}$ \\ 
$p_{13} = \frac{2-\sqrt{3}}{14}$ 
& $e^{i\nu_{13}} = \omega_{24}$ \\ 
$p_{14} = \frac{\sqrt{3}}{14}$
& $e^{i\nu_{14}} = i \omega_{24} \left( P_5\right)^\frac{1}{4}$ \\ 
$p_{15} = \frac{4\sqrt{3}-3}{42}$ 
& $e^{i\nu_{15}} = \omega_{12}^5 (P_5)^\frac{1}{3} (Q_{13})^\frac{1}{2}
\left( -\frac{P_{13}}{\omega_3Q_{13}}\right)^\frac{1}{6}$ \\ 
$p_{16} = \frac{7-4\sqrt{3}}{14}$ 
& $e^{i\nu_{16}} = \omega_{12}^7 (P_5)^\frac{1}{12} 
\left( - \frac{\omega_3Q_{13}}{P_{13}}\right)^\frac{1}{12}$ \\ 
$p_{17} = \frac{4\sqrt{3}-6}{21}$
& $e^{i\nu_{17}} = (P_5)^\frac{1}{6} 
\left( - \frac{\omega_3Q_{13}}{P_{13}}\right)^\frac{1}{12}$ \\  
$p_{18} = \frac{1}{7}$ 
& $e^{i\nu_{18}} = - (P_5)^\frac{1}{6} 
\left( - \frac{\omega_3Q_{13}}{P_{13}}\right)^\frac{1}{6}$ \\ 
\hline \hline
\end{tabular}
}
\label{tab:modulid}
\end{table} 

\vspace{1cm}

\section{Group theoretic background}\label{sec:groups}

\vspace{5mm}

\noindent We must now introduce the Weyl--Heisenberg group and the Clifford group. 
The latter is the unitary automorphism group of the former. Our account is intended 
to fix notation and to highlight some special features that we will make use of. All 
the dimensions we will encounter in this paper are odd prime numbers or products of 
odd prime numbers with multiplicity one, so here we consider only such choices of 
the dimension $d$. As it happens this leads to considerable simplifications, because 
arithmetic modulo $d$ plays a large role, and in prime dimensions the set of integers 
modulo $d$ is a field, not just a ring. We will be able to deal with square-free odd 
dimensions by using the direct product structure of the relevant groups. For more 
complete accounts we refer elsewhere \cite{Marcus, Scott, DMA}.  

Keeping these limitations in mind, let us start. The Weyl--Heisenberg group acting 
on ${\mathbb C}^d$ is generated by the unitary clock and shift operators $Z$ and $X$. 
In the basis where $Z$ is diagonal they are represented by 

\begin{equation} Z|r\rangle = \omega_d|r\rangle \ , \hspace{8mm} 
X|r\rangle = |r + 1\rangle \ . \end{equation}

\noindent This effectively defines the standard basis. Its basis vectors are 
labelled by integers modulo $d$. For any $d$ we define the primitive $d$th 
root of unity 

\begin{equation} \omega_d = e^{\frac{2\pi i}{d}} \ . \end{equation}

\noindent The group generators $X$ and $Z$ are of order $d$. The group itself is 
of order $d^3$, but up to phase factors there are only $d^2$ group elements. They are 
conveniently represented by the $d^2$ displacement operators 

\begin{equation} D_{\bf p} =  D_{i,j} = \tau^{ij}X^iZ^j \ , \hspace{8mm} 
\tau = - e^{\frac{\pi i}{d}} \ , \hspace{8mm} \tau^2 = \omega_d \ . \end{equation}

\noindent Here $i, j$ are integers modulo $d$, and we think of ${\bf p}$ as a two 
component vector with entries $i$ and $j$. An orbit of the group is obtained by 
choosing a fiducial unit vector $|\Psi_{\bf 0}\rangle$ and acting on it with all the 
displacement operators. The orbit is a SIC if and only if  

\begin{equation} |\langle \Psi_{\bf 0}|D_{\bf p}|\Psi_{\bf 0}\rangle |^2 = 
\frac{1}{d+1} \label{overlaps} \end{equation}

\noindent for all ${\bf p} \neq {\bf 0}$. This is a highly non-trivial set of real 
quartic polynomial equations for the components of the fiducial vector. The complex 
quantities $\langle \Psi_{\bf 0}|D_{\bf p}|\Psi_{\bf 0}\rangle$ are known as SICs 
overlaps (or as radar ambiguity functions, for those who like to keep the engineering 
background in mind). 
 
Zauner \cite{Zauner} and Grassl \cite{Grassl6} drew attention to the group containing 
all unitary operators that take the Weyl--Heisenberg group into itself under conjugation. 
This includes the symplectic group defined  by unimodular two-by-two 
matrices with entries that are integers modulo $d$. With the representation of the 
Weyl--Heisenberg group fixed as above, the associated unitary representation of 
the symplectic group $SL(2, {\mathbb Z}/d{\mathbb Z})$ for odd prime $d$ is 

\begin{equation} F = \left( \begin{array}{cc} \alpha & \beta \\ \gamma & \delta \end{array} 
\right) \hspace{1mm} \rightarrow \hspace{1mm} \left\{ \begin{array}{ll}
U_F = \frac{e^{i\theta}}{\sqrt{d}}\sum_{r,s} 
\omega^{2^{-1}\beta^{-1}(\delta r^2 - 2rs + \alpha s^2)} 
|r\rangle \langle s| &  \beta \neq 0 \\
\\
U_F =  \mbox{{\tiny $\genfrac{(}{)}{}{0}{\alpha}{d}$}}  
\sum_s\omega^{2^{-1}\alpha \gamma s^2}|\alpha s\rangle \langle s| &  
\beta = 0 \ . \end{array} \right. 
\label{1} \end{equation}

\noindent Here  {\tiny $\genfrac{(}{)}{}{0}{\alpha}{d}$} $= \pm 1$ is the Legendre symbol; recall that we have assumed 
that $2$ has the multiplicative inverse $2^{-1}$ modulo $d$. One finds that 

\begin{equation} U_FD_{\bf p}U_F^{-1} = D_{F{\bf p}} \ . \label{conjugation} \end{equation}

\noindent A symplectic matrix is represented by a monomial unitary matrix if and only if 
$\beta = 0$, 
as seen in eq.\ (\ref{1}). This happens because such matrices transforms the maximal 
abelian subgroup generated by the diagonal operator $Z = D_{0,1}$ into itself. To obtain 
a faithful representation the phase factors in eq. (\ref{1}) are chosen to be    

\begin{equation} e^{i\theta} = - \frac{1}{i^{\frac{d+3}{2}}} 
\mbox{{\tiny $\genfrac{(}{)}{}{0}{-\beta}{d}$}} = \left\{ 
\begin{array}{lll} (-1)^k\mbox{{\tiny $\genfrac{(}{)}{}{0}{-\beta}{d}$}} 
& \mbox{if} & d = 4k+1 \\ \\ (-1)^{k+1}i\mbox{{\tiny $\genfrac{(}{)}{}{0}{-\beta}{d}$}} & 
\mbox{if} & d = 4k+3 \ . \end{array} \right. \end{equation}

\noindent The faithful representation is also referred to as the metaplectic 
representation, and sometimes as the Weil representation. 
The choice of phase factors ensures---via a Gauss sum---that the representation is 
in terms of numbers belonging to the cyclotomic field ${\mathbb Q}(\omega_d )$. 
Thus, taking dimension $d =5$ as an example, it holds that 

\begin{equation} \sqrt{5} = \omega_5 + \omega^4_5 - 
\omega_5^2 - \omega_5^3 \ . \end{equation}

\noindent It follows that $\sqrt{5}$ belongs to 
the cyclotomic number field ${\mathbb Q}(\omega_5)$. 

Let us now consider composite dimensions where the dimensions of the factors are 
relatively prime. In this situation the Weyl--Heisenberg group in the composite 
dimension splits as a direct product of Weyl--Heisenberg groups in the factors. So 
does the symplectic group. The details of this isomorphism consist in a judicious 
application of the Chinese remainder theorem, first worked out by David 
Gross and described elsewhere \cite{ABDF}. We need only a special case here. 
Consider dimensions of the form $d(d-2)$ for some choice of $d$. If 
$d$ is odd then $d$ and $d-2$ are relatively prime. The group structure 
dictates that we write  

\begin{equation} {\mathbb C}^{d(d-2)} = {\mathbb C}^{d-2}\otimes {\mathbb C}^{d} \ . 
\label{entan} \end{equation}

\noindent We need the explicit form of the Chinese remainder isomorphism for 
the symplectic group. In our special case, using a subscript to imply the modulus 
of the arithmetic, it is  

\begin{equation} 
\left( \begin{array}{cc} \alpha & \beta \\ \gamma & \delta \end{array} \right)_{d(d-2)} 
\sim \left( \begin{array}{cc} \alpha & \kappa^{-1} \beta \\ \kappa \gamma & \delta 
\end{array} \right)_{d-2} \times \left( \begin{array}{cc} \alpha & \kappa^{-1} \beta \\ 
\kappa \gamma & \delta \end{array} \right)_{d} \ , \end{equation}

\noindent where 

\begin{equation} \kappa = \frac{d-1}{2} \ . \end{equation}

\noindent This is an integer because $d$ is odd, and invertible as $-2\kappa = 1$ modulo 
$d$. For the case when $d-2$ is not a prime see ref. \cite{ABDF}. 

There do exist anti-unitary operators obeying eq.\ (\ref{conjugation}). They represent 
two-by-two matrices of determinant $-1$ modulo $d$, and fill out what is called the 
extended Clifford group. Unitarily or anti-unitarily equivalent SICs belong to a single 
orbit of this group.

It is a remarkable fact that every known Weyl--Heisenberg SIC contains an eigenvector 
of a symplectic unitary of order three \cite{Scott, Andrew}. However, in composite 
dimensions it is not always true that every symplectic unitary of order three gives 
rise to SICs. Using the 
characteristic equation for two-by-two matrices one can show that a symplectic matrix $F$ 
whose trace equals $-1$ must obey $F^3={\one}$. When the symplectic group is defined over a 
field the converse holds, that is if $F^3 = {\one}$ then either Tr$F = -1$ or else $F$ is 
the unit matrix. Hence symplectic matrices of order three are easy to recognize in the cases 
we consider. In every known case the order three matrices that occur in the SIC problem obey 
Tr$F = -1$. If $d >3$ is an odd prime there is a unique conjugacy class of order three 
matrices, consisting of all matrices with trace $-1$, and it contains the representative 

\begin{equation} F_Z = \left( \begin{array}{cc} 0 & -1 \\ 1 & -1 \end{array} \right) 
\ . \label{Zaunermatris} \end{equation}

\noindent Matrices in this conjugacy class are known as Zauner matrices, and the 
eigenspaces of the corresponding unitaries as Zauner subspaces.  The case $d = 3$ 
is special because $-1 = 2$ modulo 3, and when the trace equals $2$ there are always 
three distinct conjugacy classes. As representatives we may choose ${\one}$, $F_Z$, and 
$F_Z^2$, or more conveniently 

\begin{equation} \left( \begin{array}{cc} 1 & 0 \\ 0 & 1 \end{array} \right) \ , 
\hspace{8mm} \left( \begin{array}{cc} 1 & 0 \\ 1 & 1 \end{array} \right) \ , \hspace{8mm} 
\left( \begin{array}{rr} 1 & 0 \\ -1 & 1 \end{array} \right) \ . \end{equation}

\noindent The point is that, according to eq. (\ref{1}), these matrices are 
represented by monomial unitary matrices. The ambiguity in dimension three  
propagates through the direct product when the Chinese remainder theorem is 
applied, so that when $d = 3\cdot (3k+1)$ or $d = 3\cdot (3k+2)$ there exist more 
than one conjugacy class of order three symplectic matrices with trace $-1$. Moreover, 
when the dimension is divisible by 3 there exist order three Clifford unitaries not 
conjugate to any symplectic unitary. But it seems that SICs invariant under order 
three Clifford unitaries not conjugate to $U_{F_Z}$ or its square exist only if 
$d = 3\cdot (3k+1)$. In this paper the only composite dimensions we 
consider are of the form $d = 3\cdot (3k+2)$. The available 
evidence suggests that every SIC in these dimensions is equivalent to one 
invariant under the unitary corresponding to $F_Z$ \cite{Scott}. 

Sometimes a symmetry of order larger than three is observed among the SICs. 
Then the unitary symmetry group is a subgroup of the centralizer of an order three 
element. For Zauner matrices belonging to the same conjugacy class as the one given 
in eq. (\ref{Zaunermatris}) the centralizer 
is abelian. Again suppose $d$ is an odd prime. Then the centralizer within the 
symplectic group has order $d+1$ if $d = 2$ mod 3 and order $d-1$ if $d = 1$ mod 3 
\cite{DMA}. Thus it is 
large enough, or almost large enough, to define a preferred basis in ${\mathbb C}^d$. 

Published lists of exact and numerical solutions include only one SIC from each 
orbit under the extended Clifford group, and by convention the matrix $F_Z$ is 
chosen as a representative of its conjugacy class \cite{Scott, Andrew}. 
However, when $d = 1$ mod 3 it is possible to choose the representative 
of the conjugacy class so that the entire centralizer is given by diagonal 
two-by-two matrices, hence by monomial unitary matrices. This leads to considerable 
simplifications that we will make use of.  

Readers familiar with the representation theory of $SL(2,{\mathbb R})$ will recognize 
several ingredients of our discussion, such as the occurence of 
parabolic conjugacy classes when the trace $t=2$, and diagonalizable matrices 
when the trace $t$ obeys $t^2-4 >0$. In the present case the inequality translates 
to the requirement that $t^2-4$ be a quadratic residue. The story would have been 
more complicated had we not restricted ourselves to dimensions that are 
products of odd prime numbers of multiplicity one. We refer to the 
literature for this \cite{Marcus, Scott}. For an in-depth study of order three 
elements of the Clifford group in the general case, see Bos and Waldron \cite{Bos}.   

\vspace{1cm}

\section{The adapted basis}\label{sec:basis}

\vspace{5mm}

\noindent In Section \ref{sec:solutions} we left the adapted basis undefined. This must now 
be remedied. It is an easy task because we view the higher dimensional Hilbert spaces as tensor 
products, 

\begin{equation} {\mathbb C}^{15} = {\mathbb C}^3\otimes {\mathbb C}^5 \ , \hspace{8mm} 
{\mathbb C}^{195} = {\mathbb C}^{13} \otimes {\mathbb C}^{15} = 
{\mathbb C}^{13} \otimes {\mathbb C}^3\otimes {\mathbb C}^5 \ ,  
\label{tensorprod} \end{equation}

\noindent and so on as we walk up the dimension ladder. The bases consist of product 
vectors, so it is enough to consider the factor spaces separately. In dimensions 13 
(a prime equal to 1 mod 3) and 3 the conjugacy class of Zauner matrices has a member 
represented by a monomial unitary matrix, so in these two dimensions 
the calculations are trivial. The calculations needed in dimension 5 are somewhat 
lengthy but still straightforward. 

Following Zauner, what we do to find a SIC in dimension 
5 is to choose any order three symplectic matrix. Then we calculate the eigenvectors 
of the corresponding unitary. Thus 

\begin{equation}
F_Z = \left( \begin{array}{rr} 0 & -1 \\ 1 & -1 \end{array} 
\right)_{5} \hspace{5mm} \Rightarrow \hspace{5mm} 
U_{Z} = \frac{-1}{\sqrt{5}}\left( \begin{array}{ccccc}
1 & 1 & 1 & 1 & 1 \\ \omega_5^3 & \omega_5^4 & 1 & \omega_5 & \omega_5^2 \\ 
\omega_5^2 & \omega_5^4 & \omega_5 & \omega_5^3 & 1 \\ 
\omega_5^2 & 1 & \omega_5^3 & \omega_5 & \omega_5^4 \\ 
\omega_5^3 & \omega_5^2 & \omega_5 & 1 & \omega_5^4 \end{array} \right) \ . 
\label{zaunerunit} \end{equation}

\noindent The Zauner matrix $F_Z$ clearly commutes with 

\begin{equation} F_P = \left( \begin{array}{rr} -1 & 0 \\ 0 & -1 \end{array} \right) 
\ . \end{equation}

\noindent The latter is represented by a monomial unitary of order two, known as the 
parity operator. The order six operator $U_{ZP}$ represents $F_ZF_P$, a generator 
of the centralizer of $F_P$, and it has a non-degenerate spectrum. Hence it defines 
a preferred basis in ${\mathbb C}^5$, adapted to the SIC because---according to a 
refined form of Zauner's conjecture---there must be a SIC fiducial that is an 
eigenvector of $U_Z$. 

After a routine calculation one finds eigenvectors that, when expressed as vectors 
relative to the standard basis, are

\begin{equation} |e_0^{(5)} \rangle = \frac{1}{\sqrt{n_0}} 
\left( \begin{array}{c} 4\omega_5-2\omega_5^4 +2\omega_3(\omega_5^3-\omega_5^4) \\ 
\omega_5^3 + \omega_5^4 + \omega_3(1 + 2\omega_5^3 + 2\omega_5^4) \\ 
4\omega_5^2 + 3\omega_5^4 +3\omega_3(\omega_5^4-1) \\ 
4\omega_5^2 + 3\omega_5^4 +3\omega_3(\omega_5^4-1) \\ 
\omega_5^3 + \omega_5^4 + \omega_3(1+2\omega_5^3 + 2\omega_5^4)
\end{array} \right) \  |e_1^{(5)}\rangle = \frac{1}{\sqrt{n_1}} 
\left( \begin{array}{c} 0 \\ 
\omega_5 - \omega_5^3 + \sqrt{5}\omega_3  \\ 
1-\omega_5^4 \\ -1+\omega_5^4 \\ 
 - \omega_5 + \omega_5^3 -\sqrt{5}\omega_3
\end{array} \right) \end{equation}

\begin{equation} |e_2^{(5)} \rangle = \frac{1}{\sqrt{n_0}} 
\left( \begin{array}{c} 4\omega_5 -2\omega_5^4 +2\omega_3(\omega_5^3-\omega_5^4) \\ 
4+ 3\omega_5^2 + 3\omega_3(\omega_5^2 - \omega_5^3) \\ 
1 + \omega_5 -\omega_3(2\omega_5^3 + 2\omega_5^4+\omega_5^2) \\ 
1 + \omega_5 -\omega_3(2\omega_5^3 + 2\omega_5^4+\omega_5^2) \\ 
4+ 3\omega_5^2 + 3\omega_3(\omega_5^2 - \omega_5^3)
\end{array} \right) \  |e_3^{(5)}\rangle = \frac{1}{\sqrt{n_1}} 
\left( \begin{array}{c} 0 \\ 
\omega_5^3 - \omega_5^2  \\ 
\omega_5 - \omega_5^4 + \sqrt{5}\omega_3^2\omega_5^2 \\ 
\omega_5^4 - \omega_5 - \sqrt{5}\omega_3^2\omega_5^2 \\ 
\omega_5^2-\omega_5^3 
\end{array} \right) \end{equation}

\begin{equation} |e_4^{(5)} \rangle = \frac{\omega_3^2-\omega_3}{15}
\left( \begin{array}{c} 4 + 3\omega_5 + 2\omega_5^2 + 6\omega_5^3 \\ 
- 1 -2\omega_5 -3 \omega_5^2 + \omega_5^3 \\ 4+3 \omega_5 +2\omega_5^2 + \omega_5^3  \\ 
4+3 \omega_5 +2\omega_5^2 + \omega_5^3 \\ - 1 -2\omega_5 -3 \omega_5^2 + \omega_5^3 
\end{array} \right) \ . \label{zauneregen} \end{equation}

\noindent We constructed the second pair of unit vectors from the first by 
applying the anti-unitary transformation representing the matrix $F = 2{\one}$, 
which is why there are only two independent normalization factors $n_0$ and $n_1$. 
They are easily calculated, but their square roots $\sqrt{n_0}$ and $\sqrt{n_1}$ do 
not belong to the cyclotomic field ${\mathbb Q}(\omega_5, \omega_3)$. We will 
have to return to this issue in Section \ref{sec:checks}.  

The basis vectors may look complicated, but they are the result of a routine 
calculation and are easily described in words. It is sometimes helpful to 
label the normalized eigenvectors with the eigenvalues of $U_Z$ and $U_P$,  

\begin{eqnarray} |e^{(5)}_0\rangle  = |\omega_3, +\rangle_5 
\ , \hspace{8mm}  |e^{(5)}_1\rangle =|\omega_3, -\rangle_5 \ , \hspace{16mm} 
\nonumber \\ \label{5abas} \\   
|e^{(5)}_2\rangle   = |\omega_3^2, +\rangle_5 \ , 
 \hspace{5mm}  |e^{(5)}_3\rangle
 =|\omega_3^2, -\rangle_5 \ , \hspace{5mm}  |e^{(5)}_4\rangle   
 =|1, +\rangle_5 \ .  \nonumber \end{eqnarray}
 
\noindent Following Zauner \cite{Zauner} 
we have chosen their overall phase factors by first ensuring that their first 
non-vanishing components are either real or purely imaginary, and afterwards we 
have multiplied them with suitable powers of $\omega_5$ with a view to simplify 
the calculations in Section 5. 

The logic behind the adapted bases in dimensions 15 and 195 is similar, but because of 
eq. (\ref{tensorprod}) the lengthy calculations are already behind us. The point is that 
the dimensions of the Hilbert spaces we will be factoring in, ${\mathbb C}^3$ and 
${\mathbb C}^{13}$, admit monomial Zauner unitaries. We are free to choose the 
representative of the conjugacy 
class of the symplectic symmetries in the higher dimension to exploit this fact. 

In dimension 15 we therefore use the Zauner matrix 

\begin{equation} F_{Z_{15}} =  \left( \begin{array}{rr} 1 & 0 \\ -1 & 1 \end{array} 
\right)_{3} \times \left( \begin{array}{rr} 0 & -1 \\ 1 & -1 \end{array} 
\right)_{5} \sim \left( \begin{array}{rr} -5 & 3 \\ -2 & 4 \end{array} 
\right)_{15} \ . \end{equation}

\noindent This choice ensures that the matrix in the dimension 3 factor is represented  
by a monomial unitary matrix. According to by now standard conventions, 
Scott and Grassl \cite{Scott} use the Zauner matrix (\ref{Zaunermatris}) 
in their tables of solutions. 
The transformation between the two representatives is 

\begin{equation} \left( \begin{array}{rr} -7 & -1 \\ -7 & 1 \end{array} 
\right)_{15} \left( \begin{array}{rr} 0 & -1 \\ 1 & -1 \end{array} \right)_{15} 
\left( \begin{array}{rr} 1 & 1 \\ 7 & -7 \end{array} 
\right)_{15} = \left( \begin{array}{rr} -5 & 3 \\ -2 & 4 \end{array} 
\right)_{15}\ . \end{equation}

\noindent We will transform numerical fiducials from the Scott and Grassl form to 
the form we prefer. We use the metaplectic representation of the symplectic group in the 
prime dimensional factors, and find that the SIC fiducials lie in the six dimensional 
Zauner subspace corresponding to eigenvalue $\omega_3^2$. 

We use the Zauner and parity unitaries to define an adapted basis in the ${\mathbb C}^3$ 
factor, namely 

\begin{equation} 
|e_0^{(3)}\rangle = \frac{1}{\sqrt{2}} 
\left( \begin{array}{c} 0 \\ 1 \\ 1 \end{array} \right) \hspace{8mm} 
|e_1^{(3)}\rangle = \frac{1}{\sqrt{2}} 
\left( \begin{array}{r} 0 \\ 1 \\ -1 \end{array} \right) \hspace{8mm} 
|e_2^{(3)}\rangle = \left( \begin{array}{c} 1 \\ 0 \\ 0 \end{array} \right) \ ,  
\end{equation}

\noindent or in alternative notation 

\begin{equation} 
|e_0^{(3)}\rangle = |\omega_3,-\rangle_3 \ , \hspace{8mm} |e_1^{(3)}\rangle = 
|\omega_3,+\rangle_3 \ , \hspace{8mm}  
|e_2^{(3)}\rangle = |1,-\rangle_3 \ . \label{3bas} \end{equation}

\noindent The square root of 2 will require attention in Section \ref{sec:checks}. 

The alternative notation is useful when we construct the adapted normalized basis 
in the Zauner subspace (eigenvalue $\omega_3^2$) of ${\mathbb C}^{15}$. It consists 
of the product vectors  

\begin{equation} \begin{array}{lll} 
|e_0^{(15)}\rangle &  = |\omega_3,-\rangle_3 \otimes |\omega_3,+\rangle_5 
& = |e_0^{(3)}\rangle |e_0^{(5)}\rangle \\ 
|e_1^{(15)}\rangle &= |\omega_3,-\rangle_3 \otimes |\omega_3,-\rangle_5 
& = |e_0^{(3)}\rangle |e_1^{(5)}\rangle \\  
|e_2^{(15)}\rangle & = |1,-\rangle_3 \otimes |\omega^2_3,+\rangle_5
& = |e_2^{(3)}\rangle |e_2^{(5)}\rangle \\ 
|e_3^{(15)}\rangle & = |1,-\rangle_3 \otimes |\omega_3^2,-\rangle_5 
& = |e_2^{(3)}\rangle |e_3^{(5)}\rangle \\ \\ 
|e_4^{(15)} \rangle & = |\omega_3,+\rangle_3 \otimes |\omega_3,-\rangle_5 
& = |e_1^{(3)}\rangle |e_0^{(5)}\rangle \\ 
|e_5^{(15)}\rangle & = |\omega_3,+\rangle_3 \otimes |\omega_3,+\rangle_5 
& = |e_1^{(3)}\rangle |e_1^{(5)}\rangle \ . \end{array} 
\label{lexicograph} \end{equation}

\noindent The labelling of the basis vectors may seem odd. We have divided them into 
two groups because fiducial 15d has an additional symmetry, implying that it lies in a 
four dimensional subspace of the Zauner subspace. The first group of basis vectors 
spans that subspace within a subspace. Within the two groups of vectors 
lexicographical ordering is being used, in a way that is manifested after the second 
equality sign.  

We now factor in ${\mathbb C}^{13}$ in order to reach dimension $195$. Fiducial 195d 
has a symmetry group of order 12 \cite{ABDF}. We choose the representative of the 
conjugacy class to ensure that it is represented by monomial unitary matrices 
in two of the factors, and obtain two commuting generators of the symmetry group in 
the form 

\begin{equation} F_Z = \left( \begin{array}{rr} 55 & 156 \\ 169 & 139 \end{array} 
\right)_{195} \sim  \left( \begin{array}{rr} 3 & 0 \\ 0 & -4 \end{array} 
\right)_{13} \times \left( \begin{array}{rr} 1 & 0 \\ -1 & 1 \end{array} 
\right)_{3}\times \left( \begin{array}{rr} 0 & -1 \\ 1 & -1 \end{array} 
\right)_{5} \ \end{equation}

\begin{equation} F_S = \left( \begin{array}{rr} 161 & 0 \\ 0 & 86 \end{array} 
\right)_{195} \sim  \left( \begin{array}{rr} 5 & 0 \\ 0 & -5 \end{array} 
\right)_{13} \times \left( \begin{array}{rr} -1 & 0 \\ 0 & -1 \end{array} 
\right)_{3}\times \left( \begin{array}{rr} 1 & 0 \\ 0 & 1 \end{array} 
\right)_{5} \ . \end{equation}

\noindent The first matrix here is a Zauner matrix, while the order four 
matrix $F_S$ is represented by a unitary operator $U_S$. Fiducial 195d 
sits in an eigenspace of $U_S$ having eigenvalue $1$. Fiducial 195b has a lower 
symmetry, and is invariant under $U_S^2$ only. 

An adapted basis in the ${\mathbb C}^{13}$ factor is easily constructed. The centralizer 
of its Zauner matrix is of order 12, and is represented by monomial matrices. From 
eq. (\ref{1}) we see that, in the unitary representation we use, the vector $|0\rangle$ 
is special so that there is a natural split 

\begin{equation} {\mathbb C}^{13} = {\mathbb C}^1 \oplus {\mathbb C}^{12} \ .  
\label{13sum} \end{equation}

\noindent Thus the centralizer is generated by a unitary matrix of the form 

\begin{equation} U_{{13}} = - \left( \begin{array}{c|c} 1 & 0 \\ \hline 0 & \Pi 
\end{array} \right) \ , \end{equation}

\noindent where $\Pi$ is a 12 by 12 permutation matrix with a non-degenerate spectrum 
consisting of 12th roots of unity. The eigenvectors are easily written down, and 
normalized, using only numbers from the cyclotomic field ${\mathbb Q}(\omega_{12})$. 
For definiteness we choose their phases so that their first non-zero entry is real. 
Incidentally things would be equally simple were we to take another step up the 
dimensional ladder, reaching dimension ${\mathbb C}^{193}\otimes {\mathbb C}^{195}$. 
Then we would encounter a unitary matrix $U_{193}$ having a form analogous to that 
of $U_{13}$.

The $d = 195$ SIC fiducials we are interested in are invariant under 

\begin{equation} U_S^2 = U_P\otimes {\one}_{15} \ , \end{equation}

\noindent where $U_P$ is the parity operator in dimension 13 \cite{Andrew, ABDF}. 
This means that only sixth roots of unity occur as eigenvalues in the relevant 
subspace of ${\mathbb C}^{13}$. It is convenient to label the eigenvectors with the 
eigenvalues of the Zauner unitary and of the order four operator $U_S$. The 
eigenvector in the ${\mathbb C}^1$ summand in eq. (\ref{13sum}) is 

\begin{equation} |e_0^{(13)}\rangle = |1,-,a\rangle \ . \end{equation}

\noindent To span the relevant subspace of ${\mathbb C}^{195}$ we will need the additional 
eigenvectors 

\begin{eqnarray} |e_1^{(13)}\rangle = |1,-,b\rangle \ , \hspace{6mm} 
|e_2^{(13)}\rangle = |\omega_2^3,+\rangle \ , \hspace{5mm} |e_3^{(13)}\rangle 
= |\omega_3,-\rangle \ , \nonumber \\ \\
|e_4^{(13)}\rangle = |1,+\rangle \ , \hspace{10mm} 
|e_5^{(13)}\rangle = |\omega_2^3,-\rangle \ , \hspace{5mm} |e_6^{(13)}\rangle 
= |\omega_3,+\rangle \ . \nonumber \end{eqnarray}

\noindent In fact all the components of these vectors are in ${\mathbb Q}(\omega_3)$. 

We are now ready to write down the adapted basis for the eigenspace in ${\mathbb C}^{195}$ 
that holds our SIC fiducials. Its dimension is 36, but fiducial 195d has a further 
symmetry and sits in a smaller subspace of dimension 19 \cite{ABDF}. We order the 
product basis vectors spanning the latter lexicographically, and arrive at

\begin{eqnarray} & |e_0\rangle = |1,-,a\rangle |\omega_3 , -\rangle |\omega_3,+\rangle 
\hspace{8mm} & |e_1\rangle = |1,-,a\rangle |\omega_3 , -\rangle |\omega_3,-\rangle 
\nonumber \\ 
&|e_2\rangle = |1,-,a\rangle |1 , -\rangle |\omega_3^2,+\rangle 
\hspace{10mm} & |e_3\rangle = |1,-,a\rangle |1 , -\rangle |\omega_3^2,-\rangle \nonumber \\
&|e_4\rangle = |1,-,b\rangle |\omega_3 , -\rangle |\omega_3,+\rangle 
\hspace{8mm} & |e_5\rangle = |1,-,b\rangle |\omega_3 , -\rangle |\omega_3,-\rangle 
\nonumber \\ 
&|e_6\rangle = |1,-,b\rangle |1 , -\rangle |\omega_3^2,+\rangle 
\hspace{10mm} &|e_7\rangle = |1,-,b\rangle |1 , -\rangle |\omega_3^2,-\rangle \nonumber \\
&|e_8\rangle = |\omega_3^2,+\rangle |\omega_3 , +\rangle |\omega_3^2,+\rangle 
\hspace{9mm} & |e_9\rangle = |\omega_3^2,+\rangle |\omega_3 , +\rangle |\omega_3^2,-\rangle 
\label{storbas} \\ 
&|e_{10}\rangle = |\omega_3,-\rangle |\omega_3 , -\rangle |1,+\rangle 
\hspace{10mm} & |e_{11}\rangle = |\omega_3,-\rangle |1 , -\rangle |\omega_3,+\rangle 
\nonumber \\ 
&|e_{12}\rangle = |\omega_3,-\rangle |1 , -\rangle |\omega_3,-\rangle \hspace{10mm} 
&|e_{13}\rangle = |1,+\rangle |\omega_3 , +\rangle |\omega_3,+\rangle \nonumber \\
&|e_{14}\rangle = |1,+\rangle |\omega_3 , +\rangle |\omega_3,-\rangle \hspace{10mm} 
&|e_{15}\rangle = |\omega_3^2,-\rangle |\omega_3 , -\rangle |\omega_3^2,+\rangle \nonumber \\ 
&|e_{16}\rangle = |\omega_3^2,-\rangle |\omega_3 , -\rangle |\omega_3^2,-\rangle \hspace{8mm} 
&|e_{17}\rangle = |\omega_3^2,-\rangle |1 , -\rangle |1,+\rangle \nonumber \\
&|e_{18}\rangle = |\omega_3,+\rangle |\omega_3 , +\rangle |1,+\rangle \ .  \hspace{8mm}   
&  \  \nonumber 
\end{eqnarray} 

\noindent Fiducial 195d can be expanded in terms of these. An additional set of 17 basis 
vectors is needed to obtain a basis in which fiducial 195b can be expanded. Again we order 
the product vectors lexicographically within this set, as explained below eqs. 
(\ref{lexicograph}).  

This concludes the construction of the adapted basis. The calculations needed 
to construct it are admittedly somewhat lengthy, but they are entirely 
straightforward except that a choice of phase factor has to be made for each vector. 
The choices we made were partly taken from Zauner \cite{Zauner}, and partly determined 
by trial and error in the course of the calculations described in the next section. 
In Section \ref{sec:checks} we must come back to discuss the seemingly innocuous 
normalization factors $\sqrt{n_0}, \sqrt{n_1}$, and $\sqrt{2}$.  

\vspace{1cm}

\section{Converting numerical SICs to exact SICs}\label{sec:numerics}

\vspace{5mm}

\noindent Although in a way this is the key section it is a brief one. This is so 
because, with the groundwork laid, the procedure by means of which candidate exact 
solutions are obtained is very simple. It works as well as it does because we carefully 
selected the cases we looked at---and also because the phase factors in the adapted 
basis vectors were carefully chosen.

We start with a numerical SIC with 150 digits precision in the standard basis 
(kindly supplied by Andrew Scott, in the case of $d = 195$). We apply 
unitary transformations and 
Chinese remaindering to it, so that it takes the form in eq.\ (\ref{15fid}) 
with respect to the adapted basis defined in Section 4. In this basis we calculate 
the squares $p_r$ of the absolute values and suitable powers of the relative 
phases $e^{i\nu_r}$. Then we apply the `RootApproximant' command in Mathematica 
to these decimal numbers. This command implements an integer relation algorithm 
which turns out to be surprisingly effective. For the absolute values squared the answer 
comes out directly. For suitable powers of the phases Mathematica typically returns a 
fourth order polynomial equation which we solve. The results are as stated in Table 
\ref{tab:modulidb} for fiducials 15d and 15b, Table \ref{tab:modulid} for 
fiducial 195d, and Table \ref{tab:modulib} for fiducial 195b. 

To test the result we lower the precision in steps of 5 digits, in order to 
find out the minimum precision needed. Worst case results are given in Table 
\ref{tab:precision}. Finally the SIC condition is tested using 1000 digits 
precision. 

In dimension 15 there exists a total of four inequivalent fiducials, belonging 
to different orbits of the extended Clifford group. Table \ref{tab:precision} 
includes some information about how our procedure fares when we 
apply it to the fiducials 15a and 15c, which lie in a larger number 
field than 15b (and 15d) \cite{ACFW}. We did establish that for 15a and 15c the 
absolute values squared of the components are not expressible in terms of numbers 
from the base field ${\mathbb Q}(\sqrt{3})$. 

\begin{table}[h]
\caption{{\small Minimum precision needed for the calculation. For 15ac we did 
not pursue the only partially successful calculation to higher than 300 digits.}}
  \smallskip \smallskip
{\renewcommand{\arraystretch}{1.6}
\begin{tabular}
{|c|ccccc|}\hline \hline
Fiducial & 15d & 15b & 15ac & 195d & 195b \\ \hline
Moduli & 10 & 10 & $>300$ & 20 & 20 \\ 
Phases & 45 & 30 & ? & 65 & 90 \\ 
\hline \hline
\end{tabular}
}
\label{tab:precision}
\end{table} 

\vspace{1cm}

\section{Ray class fields and the SIC condition}\label{sec:checks}

\vspace{5mm}

\noindent It remains to carry out an exact check of the SIC condition. In trying to 
do so one quickly realizes that the adapted basis is not well suited to describe the 
action of the Weyl--Heisenberg displacement operators. Moreover, while the 
form taken by the components of the fiducials with respect to the orthonormal 
basis (see Tables \ref{tab:modulidb}, \ref{tab:modulid}, and \ref{tab:modulib}) is 
interesting, it 
also poses a problem. The absolute values squared lie in the base field 
${\mathbb Q}(\sqrt{3})$, but one can check that the absolute values themselves, the 
$\sqrt{p_r}$, do not belong to the minimal field predicted by the ray class 
hypothesis. Neither do the phases $e^{i\nu_r}$, nor do the actual components 
$\sqrt{p_r}e^{i\nu_r}$ because of the normalizing factors $\sqrt{n_0}, \sqrt{n_1},
 \sqrt{2}$ that were used to normalize the adapted basis. Fortunately this is easy 
to remedy once one has understood the relevant ray class field. The field is a 
vector space over the rationals, with a dimension equal to its degree, and we 
can adapt a basis for that vector space to the SIC vectors we have found. 
 
The largest ray class field to be considered here is that with conductor $d = 195$. 
It contains ray class fields with conductors 3, 5, and 13 as subfields. 
(We remind the reader that we are all the the time referring to ray class fields 
that are ramified at both infinite places. Readers unfamiliar with number theory may 
prefer to ignore this point.) The computer program Magma provides us with suggestions 
for suitable generators of these fields, which we then bring to the desired form. This 
will provide the adapted basis for the number fields.  

We have carried out the analysis for the fiducial vectors $5a, 15db, 195db$. Here 
we describe it for the case $195d$. Then the field has degree $2^8\cdot 3^3$, so 
we will need eight quadratic and three cubic generators. It turns out that the 
SIC fiducial vectors we have written down can be expressed in terms of a subspace of 
the field. The Ray Class Hypothesis says that the ratios between the components of 
the fiducial vector together with the roots of unity needed to create the SIC 
generate the ray class field, 
and this will still hold in our case. A partial decoupling of the cyclotomic field 
from the SIC fiducial has been noticed before \cite{GS}, and has been used to great 
effect in unpublished work by Markus Grassl \cite{SGprivate}. 

For the ray class field with conductor $d = 5$ we use the generators 

\begin{eqnarray} a = \sqrt{3} \hspace{10mm} i = \sqrt{-1} \hspace{10mm} 
r_1 = \sqrt{5} \hspace{10mm} b_1 = \sqrt{(5-r_1)/2} \nonumber \\ \\ 
 \hspace{10mm} b_2 =\sqrt{((10a-20)r_1+10a+20)b_1+20ar_1-100a} \hspace{5mm} \ . \nonumber \end{eqnarray}
 
\noindent The first four of these generators generate the cyclotomic subfield 

\begin{equation} {\mathbb Q}(\omega_{60}) ={\mathbb Q} (a,i,r_1,b_1) \ . \end{equation}

\noindent We also introduce two quadratic generators that do not belong to the 
ray class fields of interest, namely 

\begin{equation} e_1 = \sqrt{2a} \hspace{12mm} e_2 = \sqrt{2} \ . \end{equation}

\noindent Thus equipped we can take up the issue of how to define the adapted basis. 
Before doing so we remark that the generator $e_1$ is exactly what one needs to 
construct the larger fields containing the fiducials 15b and 195b. The generator 
$e_2$ is needed for the fiducials 15a and 15c, which are considered elsewhere 
\cite{ACFW}. 

In Section \ref{sec:basis} we had to introduce square roots of numbers in the 
cyclotomic field in order to normalize the basis vectors $|e_i^{(5)}\rangle$ 
and $|e_i^{(3)}\rangle$. This takes us out of the ray class 
field. It is here that the generators $e_1$ and $e_2$ enter. We simply write  

\begin{eqnarray} |e_0^{(5)}\rangle = e_2|f_0^{(5)}\rangle \ , \hspace{5mm} 
|e_1^{(5)}\rangle = b_2e_1e_2|f_1^{(5)}\rangle \ , \hspace{5mm} 
|e_2^{(5)}\rangle = e_2|f_3^{(5)}\rangle \ , \nonumber \\ \\ 
|e_3^{(5)}\rangle = b_2e_1e_2|f_3^{(5)}\rangle \ , \hspace{10mm} 
|e_4^{(5)}\rangle = |f_4^{(5)}\rangle \ , \hspace{16mm} \nonumber \end{eqnarray}

\begin{equation} |e_0^{(3)}\rangle = e_2|f_0^{(3)}\rangle \ , \hspace{7mm} 
|e_1^{(3)}\rangle = e_2|f_1^{(3)}\rangle \ , \hspace{7mm} 
|e_2^{(3)}\rangle = |f_2^{(3)}\rangle \ . \end{equation}

\noindent The vectors $|f_i^{(5)}\rangle$ and $|f_i^{(3)}\rangle$ are not unit vectors 
in general, but they have all their components in ${\mathbb Q}(\omega_{60})$ relative to 
the standard basis. The basis vectors $|e_i^{(13)}\rangle$ have components in 
${\mathbb Q}(\omega_3)$ as they stand, and need no polish.  

This attended to, we move on to the ray class field with conductor 13. It has degree 
$2^4\cdot 3$, and is generated by $a,i,r_{13},b_{13},c_{13}$, where  

\begin{eqnarray} r_{13} = \sqrt{13} \hspace{12mm} b_{13} = \sqrt{(-3r_{13}-13)/2} 
\nonumber \\ \\ c_{13}: \ x^3-13x^2+26x - 13 \ . \hspace{8mm} \nonumber \end{eqnarray}

\noindent The cubic generator $c_{13}$ is a root of the cubic equation stated. 
These three generators enter the cyclotomic field ${\mathbb Q}(\omega_{13})$.  

To reach the field with conductor $195$ we need one more quadratic and two more cubic 
generators. We choose the latter with one eye firmly on Table \ref{tab:modulid}, and 
arrive at  

\begin{equation} b_3 = \sqrt{3-4a} \hspace{10mm} c_1 = (P_5)^\frac{1}{3} \hspace{10mm} 
c_2 = \left( - \frac{P_{13}}{\omega_3Q_{13}}\right)^\frac{1}{3} \ . \end{equation}

\noindent The generator $c_1$ is in the subfield having conductor 15. A Magma 
calculation confirms that $a,i,r_1,b_1,b_2,b_3,c_1,c_2,r_{13}, b_{13}, c_3$ 
suffice to construct a basis for the ray class field with conductor 195 and 
ramification at both infinite places. 

It is now straightforward to express the components of our SIC fiducials as linear 
combinations over the rationals of products of the generators. For 195d we give the 
result semi-explicitly in Table \ref{tab:factord}. To squeeze the result into the Table 
we had to replace factors sitting in the subfield ${\mathbb Q}(a,i,r_1)$ with a 
generic ``$X$'', but one sees by inspection that the `extra' generators $e_1$ and 
$e_2$ cancel out. So does the extra generator 

\begin{equation} e_3 = \sqrt{50a-84} \end{equation}

\noindent which is needed to split one of the components into absolute value and 
phase. Note also that we multiplied the vector with an overall factor of $\sqrt{14}$ 
before doing the analysis. This results in a harmless factor of 14 in the fiducial 
projector. 

We do not give the analysis of the remaining fiducials here, but one curious feature 
of the fiducials 15b and 195b is worth mentioning. If they are split into two, one 
part lying in the subspace of Hilbert space containing fiducials 15d respectively 195d, 
and one part lying in the orthogonal complement, then, after multiplying the fiducials 
with harmless overall factors, the first parts are found to lie 
in the ray class field. In other words, the generator $e_1$ enters only in the 
components of the orthogonal complements.

{\small
\begin{table}[h]
\caption{{\small Numbers in 195d. Numbers in ${\mathbb Q}(a,i,r_1)$ are denoted by a 
generic $X$ since they are just a little too large for the table.  
The components of the ket vectors $|i;j;k\rangle \equiv |e_i^{(13)}\rangle 
\otimes |f_j^{(3)}\rangle \otimes |f_k^{(5)}\rangle$ are in 
${\mathbb Q}(a,i,r_1,b_1)$.}}
  \smallskip
\hskip 0.2cm
{\renewcommand{\arraystretch}{1.6}
\begin{tabular}
{|c|c|r|}\hline \hline
$\sqrt{14p_i}$ & Phases $e^{i\nu_i}$ & Vectors \ \ \ \ \\
\hline			
$b_3 r_{13}X$ & 	$1$  & $2|0;0;0\rangle$ \\
$r_{13}X$ & $e_1b_1b_3X$ & $2b_2e_1|0;0;1\rangle$ \\ 
$b_3 r_{13}e_2X$ & $c_1X$ & $e_2|0;2;2\rangle$ \\ 
$b_3 r_{13} e_2e_3X$ & $e_1e_3b_1c_1X$ & $b_2e_1e_2|0;2;3\rangle$ \\
$b_3r_{13}e_2X$ & $e_2X$ & $2|1;0;0\rangle$ \\ 
$b_3r_{13} e_1e_2X$ & 
$e_2b_1X$ & $b_2e_12|1;0;1\rangle$ \\ 
$b_3 r_{13}e_2$ & $c_1X$ & $e_2|1;2;2\rangle$ \\  
$r_{13}e_2X$ & 
 $e_1b_1b_3c_1X$ & $b_2e_1e_2|1;2;3\rangle$ \\ 
$b_3r_{13}e_1X$ & 	
$e_1b_3r_{13}c_1c_2^2X$ & $2|2;1;2\rangle$ \\ 
$e_1X$ & $b_1c_1c_2^2X$ & $2b_2e_1|2;1;3\rangle$ \\ 
$e_1e_2X$ &	
$e_1b_3r_{13}c^2_1c_2X$ & $e_2|3;0;4\rangle$ \\ 
$e_1X$ &	$e_1e_2b_3r_{13}c_2X$ & $e_2|3;2;0\rangle$ \\ 
$X$ & $e_1e_2b_1b_3r_{13}c_2X$ & $b_2e_1e_2|3;2;1\rangle$ \\ 
$e_2X$ & $e_2X$ & $2|4;1;0\rangle$ \\ 
$e_1e_2X$ & $e_2b_1X$ & $2b_2e_1|4;1;1\rangle$ \\ 
$b_{3}X$ & $r_{13}c_1c_2^2X$ & $2|5;0;2\rangle$ \\ 
$X$ &	
$e_1b_1b_3r_{13}c_1c_2^2X$ & $2b_2e_1|5;0;3\rangle$ \\ 
$e_1X$ & $e_1b_3r_{13}c_1^2c_2^2X$ & $|5;2;4\rangle$ \\  
$e_2$ & $c_1^2c_2X$ & $e_2|6;1;4\rangle$ \\ 
\hline \hline
\end{tabular}
}
\label{tab:factord}
\end{table} 
}

Having reached this point we can ask Magma to calculate all the SIC overlaps. 
Then our cubic generators become somewhat of a liability. The Ray Class Hypothesis 
states that, unlike the fiducial vector, the actual overlaps belong to a smaller 
ray class field with ramification at one of the infinite places only. Effectively, 
this means that the generator $i$ drops out of the answer, provided the remaining 
generators do not depend on it for their definition. For this reason we replaced 
the Pythagorean generators $c_1$ and $c_2$ with  

\begin{equation}  \tilde{c}_{1}: \ x^3-15x -20  \hspace{12mm} 
\tilde{c}_{2}: \ x^3 - 507x - 1014a + 2535 \ . \end{equation}

\noindent The overlap calculation was done using these generators. When the 
dimension equals 195 it is time consuming, and we carried it out only for the 
fiducials 5a, 15db, and 195d. Because an order 12 symmetry is built into 
the fiducial 195d it is not necessary to calculate all the $195\cdot 195$ overlaps. A 
subset of 1604 overlaps suffices. This is still a considerable calculation, 
as is shown by the fact that the final Magma file containing the 1604 overlaps 
takes up 72 MB. All the overlaps have absolute value squared equal to $1/196$, 
which completes the proof that Table \ref{tab:modulid}, or equivalently Table 
\ref{tab:factord}, indeed gives a SIC fiducial. Moreover, all the overlaps 
belong to the smaller ray class field mentioned above. Thus the Ray Class 
Hypothesis is fully vindicated.

\vspace{1cm}

\section{Summary and discussion}\label{sec:summary}

\vspace{5mm}

\noindent We have been following a three step procedure to find exact 
expression for SIC fiducials:  

\

\noindent 1. Express a numerical SIC in an adapted basis related to the 
standard basis in a definite way.

\

\noindent 2. Use Mathematica's `RootApproximant' command to convert the 
numerical solution to an exact one.

\

\noindent 3. Check that the solution lies in the expected number field, and using 
this check the SIC condition.

\

We comment on these steps in turn. Our choice of adapted basis makes use of the 
Chinese remaindering isomorphism. Then it is enough to find bases in the 
prime dimensional factors, and in prime dimensions the basis singled out by the 
centralizer of the Zauner matrix suggests itself. It is adapted to the symmetries 
of the SIC \cite{DMA}. The calculations in this step are tedious but 
straightforward, except that a judicious choice of phase factors is made 
in order to simplify the next step. 

Remarkably, the second step of the procedure requires only very modest 
precision. In previous work numerical SICs were converted to exact ones using the 
basis independent overlap phases. This has the advantage that it works in general, 
but the disadvantage that the actual conversion has to be carried out in a much more 
sophisticated way using a precision sometimes running to several thousand digits 
\cite{ACFW}. We needed only 90 digits in the worst case, and could manage with 
less at the expense of more work. 

The exact expression obtained in the second step turns out to involve numbers 
that do not lie in the minimal field. This is partly due to the normalization 
factors needed to make the adapted basis orthonormal, and partly due to the 
enforced split of the components into absolute values and phase factors. This 
problem evaporates on inspection. It seems noteworthy that the quadratic 
extensions in question involve field generators that are necessary to describe 
those SICs that do not lie in the ray class field. For conductor $d =195$ the calculations 
are greatly facilitated by the choice of a suitably adapted basis for the ray class field. 

Our procedure has provided strikingly simple exact expressions for the SIC 
fiducials 5a, 15d, 15b, and 195d. We also have a candidate exact expression for 195b, 
with the expected number theoretic properties.  
In the normalized adapted basis the absolute values squared of all the components 
lie in the quadratic base field of the SIC, and all the phase factors are obtained 
from Pythagorean triples and similar triples constructed using the base field. 

Are there other SICs for which a basis adapted to the symmetries of a SIC 
vector will force the absolute values squared of the components into their 
basefields? 

According to the calculations presented in Section \ref{sec:numerics}, for 
fiducials 15a and 15c the answer is ``no''. Their number fields are in fact 
larger than that of 15b \cite{ACFW}, so this is perhaps not a surprise. We 
also investigated the ten SICs in dimension 53 \cite{Andrew}, which 
share the same base field with the ones considered 
here. Adapting the basis to the centralizer of the Zauner unitary in dimension 
53 we found that the answer for all of them is ``no''. An exact solution in 
dimension 53 has been found by other means \cite{SGprivate}. 

Still we conjecture that the answer to our question is ``yes'' for every dimension 
on the particular number theoretic ladder we have been climbing. One reason for 
optimism is the fact that the dimensions continue to factor into primes equal 
to one modulo three, with all multiplicities equal to one, for the first eight rungs of 
the ladder. (We do not understand why, we simply let Mathematica do the 
factorization for us.) Hence adapted bases suitable for these rungs of the ladder 
are easily constructed, but numerical solutions for the SICs  
are lacking. Thus we cannot perform the first step in our procedure. The suspicion is 
that, if we could, the second step would be easy. When the conductor is 
$d = 195\cdot 193$ the degree of the ray class field is $2^{19}\cdot 3^5$. 
If, as seems likely, a `decoupling' of the cyclotomic field ${\bf Q}(\omega_{193})$ 
takes place then it may be possible to verify the conjecture. We take 
Kopp's recent work \cite{Kopp} as an indication that there may be a passable 
road ahead. 

Concerning SICs over other base fields, the SIC fiducials 7b, 7a, and 35j can be 
presented in an adapted basis so that all absolute values squared of the components 
lie in the base field ${\mathbb Q}(\sqrt{2})$ \cite{Irina}. They 
belong to a dimension ladder starting at $d = 7$. It seems that SICs on the 
ladder starting at $d = 9$ will not have the same feature. For ladders starting 
at an even dimension the adapted basis needs modification, because we cannot 
rely on the Chinese remaindering isomorphism in the same way. A solution of 
this technical difficulty is on its way \cite{OA}.

An interesting point is that we had to go slightly beyond the 
minimal number fields in order to reach the simplest form of the solutions. 
We feel that we have seen something similar before. 
The electromagnetic field can be expressed using the minimal number of two 
degrees of freedom per spatial point, but extending the description to 
include two additional degrees of freedom permits the elegant and useful 
description in terms of the vector potential. For SICs we have shown that 
elegance can be gained by going beyond the minimal number fields. This proved 
useful too, in the sense that we obtained previously unknown solutions. We 
do not know where this observation tends. 

\vspace{2cm}

\noindent \underline{Acknowledgements}: We thank Andrew Scott for sending his numerical 
fiducials, and Gary McConnell, Konrad Szyma\'nski, Irina Dumitru, and an anonymous 
referee for suggestions 
and help. We also thank the Mathematics Department at Stockholm University 
for allowing us to use their computer. The final version of our paper was 
very much improved by discussions with Markus Grassl. Marcus Appleby acknowledges 
support from the Australian Research Council through the Centre of Excellence in 
Engineered Quantum Systems CE170100009.

\newpage

\appendix

\section{195b}\label{sec:195b}

{\small 
\begin{table}[h]
\caption{{\small Moduli squared and relative phases in 195b. This solution is placed here to 
emphasize that an exact check of the SIC condition has not been carried out for it.}}
  \smallskip
\hskip 0.6cm
{\renewcommand{\arraystretch}{1.6}
\begin{tabular}
{|ll|ll|}\hline \hline
(Moduli)$^2$ &  & Phases & \\
\hline			
$p_0 = \frac{4\sqrt{3}-3}{91}$ & $p_{19} =
					\frac{33-18\sqrt{3}}{182}$ & $e^{i\nu_0} = 1$ & 
					$e^{i\nu_{19}} = -(-i)^\frac{1}{2}$ \\
$p_1 = \frac{2-\sqrt{3}}{7}$ & 
					$p_{20} = \frac{4\sqrt{3}-3}{182}$ & 
					$e^{i\nu_1} = \omega_4^3 \left( \frac{P_5P_{13}}{Q_{13}}\right)^{\frac{1}{4}}$ & 
					 $e^{i\nu_{20}} = \omega_4^2\left( -P_5\right)^\frac{1}{4}$ \\    
$p_2 = \frac{62\sqrt{3}-105}{364}$& 
$p_{21} = \frac{12-3\sqrt{3}}{364}$ & 
$e^{i\nu_2} = \omega_{12}\left( \frac{P_5^4}{P_{13}^3Q_{13}^3}\right)^\frac{1}{12}$ & 
$e^{i\nu_{21}} = - (-i)^\frac{1}{2}$\\ 
$p_3 = \frac{5+2\sqrt{3}}{364}$ & $p_{22} = \frac{6+5\sqrt{3}}{364}$
					 & $e^{i\nu_3} = \omega_{12}^4\left( - P_5 \right)^\frac{1}{12}$ & 
					$e^{i\nu_{22}} = \omega_4^2\left( -P_5 \right)^\frac{1}{4}$ \\
$p_4 = \frac{4\sqrt{3}-3}{1092}$ & 
					$p_{23} = \frac{5\sqrt{3}-6}{42}$ & $e^{i\nu_4} = - 1$ & 
					$e^{i\nu_{23}} = \omega_{12}^4
					\left( -\frac{iP_5^4P_{13}^5Q_2}{Q_{13}^5}\right)^\frac{1}{12}$ \\   
$p_5 = \frac{11-6\sqrt{3}}{28}$ & $p_{24} = \frac{7-4\sqrt{3}}{14}$ & 
$e^{i\nu_5} = \omega_4^2 \left( \frac{P_5}{Q_{13}^2}\right)^\frac{1}{4}$ & 
$e^{i\nu_{24}} = 
\omega_{12}^3 \left( - \frac{P_5P_{13}^2Q_2}{Q_{13}^2}\right)^\frac{1}{12}$ \\
$p_6 = \frac{115\sqrt{3}-174}{546}$ & $p_{25} = \frac{2\sqrt{3}-3}{21}$ & 
$e^{i\nu_6} = \omega_{12}^3 
\left( \frac{iP_5^4Q_{241}^3}{P_{241}^3} \right)^\frac{1}{12}$ & 
$e^{i\nu_{25}} = \omega_{12}^2 
\left( \frac{P_5^2P_{13}^2Q_2}{Q_{13}^2}\right)^\frac{1}{12}$ \\ 
$p_7 = \frac{4-\sqrt{3}}{182}$ & $p_{26} = 0$ & 
$e^{i\nu_7} = \omega_{12}^{10}\left( -P_5 \right)^\frac{1}{12}$ & --- \\  
$p_8 = \frac{2-\sqrt{3}}{14}$ & 
$p_{27} = \frac{\sqrt{3}}{84}$ & 
$e^{i\nu_8} = \omega_{12}^5 \left( \frac{P_5^4P_{13}^2Q_2}{Q_{13}^2} \right)^\frac{1}{12}$ & 
$e^{i\nu_{27}} = (i)^\frac{1}{2}$ \\  
$p_9 = \frac{2\sqrt{3}-3}{14}$ & $p_{28} = \frac{10-5\sqrt{3}}{28}$ & 
$e^{i\nu_9} = \omega_{12}^2 \left(- \frac{P_5P_{13}^2Q_2}{Q_{13}^2}\right)^\frac{1}{12}$ & 
$e^{i\nu_{28}} = \omega_{4}^2\left( -P_5^3 \right)^{\frac{1}{4}}$\\  
$p_{10} = \frac{\sqrt{3}}{21}$ & 
$p_{29} = \frac{7\sqrt{3}-12}{42}$ & 
$e^{i\nu_{10}}= \omega_{12}^5 \left( - \frac{P_5^2P_{13}}{Q_2Q_{13}}\right)^\frac{1}{12}$ & 
$e^{i\nu_{29}} = \omega_{6}^4\left( iP_5^2 \right)^{\frac{1}{6}}$ \\  
$p_{11} = \frac{\sqrt{3}}{42}$ & 
$p_{30} = \frac{2-\sqrt{3}}{14}$ & 
$e^{i\nu_{11}} = \omega_{12}^4 \left( \frac{P_{13}}{Q_2Q_{13}} \right)^\frac{1}{12}$ &
$e^{i\nu_{30}} = \omega_{12}^4\left( - P_5 \right)^{\frac{1}{12}}$ \\ 
$p_{12} = \frac{2-\sqrt{3}}{14}$ & 
$p_{31} = \frac{6 - 3\sqrt{3}}{14}$ & 
$e^{i\nu_{12}} = \omega_{12}^5 \left( \frac{P_5^3P_{13}}{Q_2Q_{13}} \right)^\frac{1}{12}$ & 
$e^{i\nu_{31}} = 
\omega_{12}^7 \left( - \frac{P_5^4Q_2Q_{13}}{P_{13}} \right)^\frac{1}{12}$ \\ 
$p_{13} = \frac{1}{28}$ & 
$p_{32} = \frac{2\sqrt{3}-3}{14}$ & $e^{i\nu_{13}} = - i$ & 
$e^{i\nu_{32}} = \omega_{12}^{11} \left( \frac{P_5Q_2Q_{13}}{P_{13}} \right)^\frac{1}{12}$ \\ 
$p_{14} = \frac{2\sqrt{3}-3}{28}$ & 
$p_{33} = \frac{4\sqrt{3}-6}{21}$ & 
$e^{i\nu_{14}} = \omega_4 \left( P_5\right)^\frac{1}{4}$ & 
$e^{i\nu_{33}} = \omega_{6}^4 \left( \frac{P_5Q_2Q_{13}}{P_{13}} \right)^\frac{1}{6}$ \\ 
$p_{15} = \frac{2\sqrt{3}-3}{42}$ & 
$p_{34} = \frac{16\sqrt{3}-27}{42}$ & 
$e^{i\nu_{15}} = \omega_{12}^{11} 
\left( \frac{P_5^4Q_2Q_{13}}{P_{13}} \right)^\frac{1}{12}$ & 
 $e^{i\nu_{34}} = 
 \left( - \frac{Q_{13}^5}{P_{13}^5Q_2}\right)^\frac{1}{12}$ \\ 
$p_{16} = \frac{2-\sqrt{3}}{14}$ & 
$p_{35} = \frac{1}{14}$ & 
$e^{i\nu_{16}} = \omega_{12}^2 
\left( - \frac{P_5Q_2Q_{13}}{P_{13}} \right)^\frac{1}{12}$ & 
$e^{i\nu_{35}} = \omega_{12}^2
\left( \frac{P_5^3Q_{13}^2}{P_{13}^2Q_2}\right)^\frac{1}{12}$ \\ 
$p_{17} = \frac{2\sqrt{3}-3}{21}$ & & 
$e^{i\nu_{17}} = \omega_{12}^{4}\left( - \frac{P_5^2Q_2Q_{13}}{P_{13}} \right)^\frac{1}{12}$ 
& \\ 
$p_{18} = \frac{7-4\sqrt{3}}{7}$ & & 
$e^{i\nu_{18}} = \omega_{12}^{11} 
\left( - \frac{P_5^2Q_{13}^2}{P_{13}^2Q_2} \right)^\frac{1}{12}$ & \\ 
\hline \hline
\end{tabular}
}
\label{tab:modulib}
\end{table} 
}


\newpage

{\small
 
\begin{thebibliography}{99}

\bibitem{FHS} C. A. Fuchs, M. C. Hoan, and B. C. Stacey, {\it The SIC question: History 
and state of play}, Axioms {\bf 6} (2017) 21. 

\bibitem{Zauner} G. Zauner: {\it Quantendesigns. Grundz\"uge einer 
nichtkommutativen Designtheorie}, PhD thesis, Universit\"at Wien, 1999; also published 
as {\it Quantum designs: Foundations of a noncommutative design theory}, 
Int. J. Quant. Inf. {\bf 9} (2011) 445.

\bibitem{Renes} J. M. Renes, R. Blume-Kohout, A. J. Scott, and C. M. Caves, 
{\it Symmetric informationally complete quantum measurements}, J. Math. Phys. {\bf 45} (2004) 2171. 

\bibitem{Marcus} D. M. Appleby, {\it SIC-POVMs and the extended Clifford group}, 
J. Math. Phys. {\bf 46} (2005) 052107.

\bibitem{Scott} A. J. Scott and M. Grassl, {\it SIC-POVMs: A new computer study}, 
J. Math. Phys. {\bf 51} (2010) 042203.

\bibitem{Zhu} H. Zhu: {\it Quantum State Estimation and Symmetric Informationally 
Complete POVMs}, PhD thesis, National University of Singapore, 2012. 

\bibitem{Andrew} A. J. Scott, {\it SICs: Extending the list of solutions}, 
arXiv:1703.03993. 

\bibitem{AYAZ} D. M. Appleby, H. Yadsan-Appleby, and G. Zauner, {\it Galois automorphisms 
of symmetric measurements}, Quantum Inf. Comp. {\bf 13} (2013) 672.

\bibitem{AFMY} M. Appleby, S. Flammia, G. McConnell, and J. Yard, 
{\it Generating ray class fields of real quadratic fields via complex 
equiangular lines}, arXiv:1604.06098.

\bibitem{GS} M. Grassl and A. J. Scott, {\it Fibonacci--Lucas SIC-POVMs}, J. Math. 
Phys. {\bf 58} (2017) 122201. 

\bibitem{SGprivate} M. Grassl and A. J. Scott, private communication.

\bibitem{DMA} D. M. Appleby, {\it Properties of the extended Clifford group with 
applications to SIC-POVMs and MUBs}, arXiv:0909.5233.

\bibitem{Ivanovic} I. D. Ivanovi\'c, {\it Geometrical description of state determination}, 
J. Phys. {\bf A14} (1981) 3241.

\bibitem{ABDF} M. Appleby, I. Bengtsson, I. Dumitru, and S. Flammia, {\it Dimension 
towers of SICs. I. Aligned SICs and embedded tight frames}, J. Math. Phys. {\bf 58} 
(2017) 122201. 

\bibitem{Kopp} G. S. Kopp, {\it SIC-POVMs and the Stark conjectures}, arXiv:1807.05877. 

\bibitem{ACFW} M. Appleby, T.-Y. Chien, S. Flammia, and S. Waldron, 
{\it Constructing exact symmetric informationally complete measurements 
from numerical solutions}, J. Phys. {\bf A51} (2018) 165302. 

\bibitem{HW} G. H. Hardy and E. M. Wright: {\it An Introduction to the Theory 
of Numbers}, 4th ed., Clarendon, Oxford 1960.

\bibitem{Grassl6} M. Grassl, {\it On SIC-POVMs and MUBs in dimension 6}, 
arXiv:quant-ph/0406175. 

\bibitem{monomial} D.~M.~Appleby, I.~Bengtsson, S.~Brierley, M.~Grassl, D.~Gross, and 
J.-\AA.~Larsson, \textit{The monomial representations of the Clifford group}, Quantum Inf.\ 
Comp.\ \textbf{12} (2012) 0404.

\bibitem{Bos} L. Bos and S. Waldron, {\it SICs and the elements of order three in 
the Clifford group}, J. Phys. {\bf A52} (2019) 105301. 

\bibitem{Irina} I. Dumitru, private communication. 

\bibitem{OA} O. Andersson and I. Dumitru, {\it Aligned SICs and embedded tight 
frames in even dimensions}, arXiv:1905.09737.


\end{thebibliography}
}
\end{document}